# Domain structures and the transitional state of ordering in long-period stacking ordered (LPSO) structures in Mg-Al-Gd alloy

Xin-Fu Gu[a, b*], Tadashi Furuhara[b], Leng Chen[a], Ping Yang[a]

[a] School of Materials Science and Engineering, University of Science and Technology Beijing, Beijing, 100083, China

[b] Institute for Materials Research, Tohoku University, Sendai, 980-8577, Japan

*Corresponding author: Xin-Fu Gu, xinfugu@gmail.com

**Abstract**

The transitional state of ordering in the long-period stacking ordered (LPSO) structure in Mg-Al-Gd alloy has been investigated by scanning transmission electron microscopy. It is found that the domains are the origin of observed transitional state. Three new patterns of clusters at this state can be fully explained by the superimposition of $L1_2$ type clusters located in different domains. The domains are further verified by different tilted views. Growth of preferred domains during ageing process will result in final ordered LPSO structure. Based on the idea of domains, less ordered LPSO structure in other dilute alloys can be also understood.

Long-period stacking ordered (LPSO) structures are contributed to considerable improvement of mechanical properties at both room and elevated temperature in various magnesium alloys [1-7]. In general, the LPSO structures can be treated as lamellar structures consisting of solute enriched stacking faults (SFs) and (0001) Mg layers in-between [8, 9]. Therefore, the four-layer height SF with fcc structure are considered as building blocks or structural unit (SU, used hereafter) in the LPSO structures. The solute atoms in the SFs would form a superstructure of ordered distributed $L1_2$ type clusters and the clusters are often in an ideal dimension of $6\times\{2\bar{1}\bar{1}0\}_\alpha$ [10-12], where the subscript α indicates the directions from Mg matrix.

The formation of LPSO structures or SUs involves a change in both composition and stacking sequence [8]. A hierarchical transformation mechanism has been proposed based on the in-situ observations by synchrotron radiation measurements [13]. Firstly, the clusters of solute atoms randomly form in Mg matrix, which may have $D0_{19}$ type structures [14]. Then, the clusters will be spatially rearranged due to the interaction between them. The SFs subsequently form passing through the clusters and the LPSO structures will be finally generated. However, the possible evolution of the clusters in the LPSO structures has not been involved in this model. On contrary, an Order-Disorder transition mechanism is proposed for the formation of LPSO structures in Mg-Al-Gd alloy [15]. Al and Gd atoms firstly enrich in four consecutive close-packed planes in the matrix prior to SF, similar to other observations [16, 17]. Later, the SF forms in the solute enriched region and then $L1_2$ type clusters ($Al_6Gd_8$) emerge. The LPSO structure is thickened by the enrichment of solute atoms in neighboring layers. In the late stage of growth or after long time ageing, in-plane ordering of $L1_2$ type clusters and changing of stacking positions between SUs in the LPSO structures have been observed [15]. The most favorable ordered structure in different types of LPSO structures has been suggested in the pioneering works by Kishida *et al.* [12, 15]. Though the ordering of clusters in the LPSO structures is recognized [11, 12, 15, 17-19], the mechanism for ordering of clusters in the SUs and the evolution of stacking positions between SUs are less known, and is needed to be further explored.

Capturing and investigating the transitional state during the transformation is inevitable for such purposes. Moreover, the LPSO structures in Mg-Al-Gd (or Y) alloys could be well ordered after prolonged annealing [18, 20]. Therefore, these alloy systems are well candidates for studying the transitional state of ordering in the LPSO structures. In a recent work, Zhang *et al.* claimed that at least three new types of metastable building clusters enriched of Al and Y atoms exist in the as-casted Mg-Al-Y alloy in addition to the well-known $L1_2$ ($Al_6Y_8$) cluster based on their observations by high angle annular dark field (HAADF) scanning transmission electron microscopy (STEM) [20]. These three metastable building clusters will be transformed to the stable $L1_2$ cluster after further annealing process. Comparatively, an alternative mechanism will be proposed in this work to explain the new patterns of clusters, and this proposal will be verified by atomic resolved images by HAADF-STEM. In addition, the ordering phenomenon in the LPSO structures will be explained thereby.

The as-casted $Mg_{92}Al_3Gd_5$ alloy (at.% default) was isothermal held at 500°C for 4 hours. The structure of the SU in a dilute $Mg_{97}Zn_1Gd_2$ alloy, which is generated by ageing of a pre-deformed sample, was used to compare with that in the Mg-Al-Gd alloy. The atomic structures of SUs or LPSO structures were observed by Cs-corrected Titan G2 60-300 (300kV, FEI). The procedure to prepare the TEM sample and the conditions for TEM observation are the same as in our previous work [21].

The ideal ordered structure in the four-layer height fcc SU is examined first. Fig. 1(a) shows an in-plane view of the superstructure of $L1_2$ type clusters ($Al_6Gd_8$) in a SU viewed along $[0001]_\alpha$ direction. The solute atoms are indicated by different colored symbols and the Mg atoms are omitted for simplicity. An $L1_2$ type cluster is separately shown in Fig. 1(b). The solute atoms in "ABCA" stacking layers are denoted by circles, downward-pointing triangle, upward-pointing triangle, and circles, respectively. The <111> view of an $L1_2$ structure will form a cluster in Fig. 1(a), the neighboring clusters in a rhombic lattice are separated by $2\sqrt{3}a$ along $<10\bar{1}0>_\alpha$ directions, where *a* is the lattice parameter of Mg matrix. Fig. 1(c) and (d) show the $<2\bar{1}\bar{1}0>_\alpha$ and $<10\bar{1}0>_\alpha$ view of the SU, respectively. A corresponding HAADF-STEM image in the Mg-Al-Gd alloy is shown at the right side of each atomic model for further reference (same for the rest). The bright dots in the images indicate the locations of Gd enriched columns according to the Z-contrast principle [22]. A $L1_2$ cluster is indicated in a dashed box in Fig. 1(d), and the ideal pattern with ideally distributed $L1_2$ clusters in the SU, such as in Fig. 1(d), is named as "$P_0$" type pattern.

Fig. 2 shows an atomic resolved HAADF-STEM image at the zone axis of $[10\bar{1}0]_\alpha$ for the sample aged at 500°C for 4 h. More images and diffraction patterns can be found in the Supplementary materials. Since the SUs are enriched with heavy Gd atoms [15, 23], and it is easy to identify the four-layer height SUs by bright contrasts in the HAADF-STEM image. The SUs are separated by two layers of Mg atoms, thus this structure is called 18R type LPSO structure (Ramsdell's notation) in a not strict way, as will be explained latter. In Fig. 2, the ordered structure of $L1_2$ type clusters ("$P_0$" type pattern) can be found in the SUs except the locations indicated by the rectangles. Within these rectangles, there are other three new type patterns of clusters as denoted by $P_1$, $P_2$ and $P_3$, which are frequently observed in the transitional state during ordering of the LPSO structures. Furthermore, according to the diffraction patterns in Fig. S1(d) or S2(d), the fundamental $(\bar{1}2\bar{1}0)$ is divided by five strips/lines indicated by the arrows, which indicates the average in-plane order is 6M, instead of 7M as recently reported in Mg-Ni-Y alloy by Yamashita *et al.* [24]. Furthermore, the distance between $L1_2$ clusters in $P_0$ type patterns can be directly measured in the HAADF-STEM image as in Fig. S2 and again the 6M-type superstructure can be found.

These three new patterns in Fig. 2 could be explained by the superimposition of domain structures in the SUs. If a part of the ordered clusters in a SU is translated with respect to their original position, a domain structure might be generated depending on the translational vectors. As inspired by previous work [12], the possible translational positions/vectors could be classified into four types as indicated by different symbols in a rhombic unit cell in Fig. 3(a), where the rhombic unit cell is the unit cell of the superstructure in the SU in Fig. 1(a). The nodes of the grid in Fig. 3(a) are the atomic positions in stacking layer A, i.e. the outer layer of the SU. The original position for Gd atoms in layer A is denoted as $t_0$ position, and the translation to its equivalent positions will not generate domain boundaries. Therefore, there are three crystallographic nonequivalent positions to generate a domain structure, i.e. $t_1$, $t_2$ and $t_3$, as shown in Fig. 3(a). Since the focal depth during HAADF-STEM imaging depends on the convergent angle, it is estimated to be about 14 nm for the convergent semi-angle 18 mrad employed in this study, thus the overlap of domain structures is possible during imaging. Due to the overlap of the domain structures, the $<2\bar{1}\bar{1}0>_\alpha$ and $<10\bar{1}0>_\alpha$ view of the SU will be different from the standard pattern "$P_0$" in Fig. 1(c-d). Fig. 3(b) shows an example of domain structure with a translation of $t = t_3 = <a>$, where the angle bracket means equivalent crystalline directions of the basal vector *a*. This kind of translation will result in two types of patterns when viewed along $<2\bar{1}\bar{1}0>_\alpha$, because the vector $t_3$ has two kinds of projection length when the translational vector is projected along different $<2\bar{1}\bar{1}0>_\alpha$ directions, i.e. 0 or $\sqrt{3}a/2$. Similarly, two types of patterns along $<10\bar{1}0>_\alpha$ view direction are also possible, i.e. "$P_3$" and "$P_1$" patterns in Fig. 3(c-d), since the translational vector $t_3$ has the projection length of $a/2$ or $a$ when the translational vector is projected along $<10\bar{1}0>_\alpha$ directions. Therefore, the formation of "P3" and "P1" type patterns in $<10\bar{1}0>_\alpha$ view could be caused by the overlap of the L1$_2$ clusters from two domains with a relative translational vector of $t_3$. In the same way, "$P_2$" type pattern as shown in Fig. 3(e) can be formed with a relative translational vector $t_2$. Table 1 summarizes the possible $<10\bar{1}0>_\alpha$ views of a SU containing a domain structure with different translational vectors $t_1$ to $t_3$, and the possible observed patterns $P_0$~$P_3$ is marked with "√" in the table, while "×" means the corresponding pattern will not be observed for the translational vector. It appears all of the three new patterns $P_1$~$P_3$ observed in our study could be explained by overlapping of the L1$_2$ clusters from two domain structures. In Fig. 2, the variation of contrast in these patterns may be due to the sample volume of different domains during imaging. It is also noted that different local ordered structure at the domain boundaries may form due to the translations between different 6M superstructure in domains. For example, in $P_3$ type patterns in Figure 3(c), the inter-cluster spacing may form locally 7M ordered structure as observed in [24]. Nevertheless, the domain structures in the LPSO structure have been observed in high alloyed $Mg_{85}Zn_6Y_9$ alloy by scanning tunneling microscopy [25, 26], but its link to the ordering process of the LPSO structures has not been revealed. Moreover, the new cluster patterns observed by HAADF-STEM have not been well rationalized, though the application of HAADF-STEM technique to investigate the LPSO structure is commonly used. If several domains are overlapped with each other, more complex ordered patterns could be expected. It should be noted that the present analysis is pure geometry, and the actual relaxed structures may vary a little from the geometrical models. Three new patterns observed in this study are similar to previous observations in Mg-Al-Y alloy, where the patterns are proposed as three new building clusters in addition to $L1_2$ cluster [20]. Comparing to previous work, an alternative mechanism due to the presence of domain structures in the SUs is proposed in our work and present analysis is self-consistent.

Furthermore, we have confirmed the domain structures with different tilted views of the SU by HAADF-STEM. The LPSO structure or SU usually has a large aspect ratio, thus it is hard to find the same position after tilting the sample. However, there are growth ledges (steps) during growth of the LPSO structures, and these steps can be used as a mark to find the same position during the tilting trails. Fig. 4(a-b) shows such an example. Fig. 4(a) and (b) show the LPSO structure viewed at $[2\bar{1}\bar{1}0]_\alpha$ and $[10\bar{1}0]_\alpha$ zone axis, respectively. With the aid of the growth step as a marker, nearly the same positions could be located in Fig. 4(a-b). In the area marked by the dashed box, the ordered pattern in Fig. 4(b) is "$P_1$" type, and the possible translation position for the domain structure would be $t_1$ or $t_3$ according to Table 1. Combined with Fig. 4(a), the translational vector could be further narrowed to be $t_3$, since the $<2\bar{1}\bar{1}0>_\alpha$ view of $t_2$ domain is the same as Fig. 1(c) and different from Fig. 4(a). It shows that the translational vector of a domain structure can be determined by different tilted experiments, and the displacement between the clusters can be used to constrain the possible translational vectors of a domain structure. Fig. 4(c) directly shows an $[0001]_\alpha$ view of the domain structures in a SU in as-casted state where well-separated SU can be observed [23]. The domain structure is identified according to the periodicity of the clusters in the image with the criterion that the clusters in one domain should be regular/ordered distributed. The distributions of the clusters in three areas are indicated by grid lattices. As is noted, these three lattice grids deviate from each other and form the domains of clusters in local areas. Accordingly, the size of the domain structures is around several nanometers. The periodicity of the clusters within the domains is consistent with distribution in Fig. 1(a). These domain structures may be formed during ordering of the planar segregation. Therefore, the overlap of domain structures would result in the new patterns often found by HAADF-STEM during the ordering process of the LPSO structures, such as in [6, 11-13].

The evolution from $P_1 \sim P_3$ patterns to $P_0$ type pattern could be caused by the growth of the domains and the elimination of the domain boundaries during prolonged ageing. The clusters in the domain boundaries usually deviate from their preferred inter-cluster distance and coordinate numbers, thus it will cause the increment of system energy [26]. Therefore, the elimination of domain boundaries is a nature process to lower the system energy. Moreover, it is noted in Fig 2 that there are different and irregular displacements between neighboring SUs as indicated by the circles, and this is why the structure in this figure is not an ideal ordered 18R type LPSO structure in a strict speaking. According to previous research [15] or Figures S1-S2 in the supplementary material, the LPSO structure will transform to be fully ordered structure with the same displacement at the late stage of growth. The ordering between the SUs could also be explained by growth of preferred domain structures, which is energetically preferred in adjacent to neighboring stacked SUs [12].

A less ordered in-plane structure of the SU in a dilute Mg-Zn-Gd alloy is shown in Fig. 4(d) for comparison, and the domain structure is small or sparsely distributed compared to that in Mg-Al-Gd alloy in Fig. 4(c). Therefore, the overlap of many domains will cause the LPSO structure less ordered, which may be one of the reasons why the LPSO structure in some alloys is not well ordered except for high alloyed ones [7, 11, 16, 18].

In summary, the idea of domain structures in the SU has been proposed to rationalize the transitional state observed in the LPSO structures by HAADF-STEM. Three new patterns of clusters observed in $<10\bar{1}0>_\alpha$ views have been fully explained by the overlap of domain structures with three crystallographic nonequivalent translational vectors. The translational vector for a domain structure can be determined by the atomic views at different zone axes. Ordering in and/or between the SUs in a LPSO structure is preceded by the growth of preferred domain structures. The concept of domain structure can be also applied to rationalize the less ordered LPSO structures observed in dilute alloys.

## Acknowledgement

This work was supported by the Ministry of Education, Culture, Sports, Science and Technology, Japan [Grant No. 23109006]; and Beijing Municipal Natural Science Foundation, China [Grant No. 2192035]. The Mg-Al-Gd alloy used in this study was provided by Kumamoto University. The authors also thank to Mr. Y. Hayasaka (Electron Microscopy Center, Tohoku University) for technical supports.

Tables

Table 1. The possible cluster patterns ($P_0$~$P_3$) observed along $<10\bar{1}0>_\alpha$ zone axes due to the superimposition of domains with its translation vector ***t*** defined in Figure 3a. The symbol '√' indicates corresponding cluster pattern in the column would be observed, while '×' indicates the cluster pattern will be absent.

| ***t*** | $P_0$ | $P_1$ | $P_2$ | $P_3$ |
|---|---|---|---|---|
| $t_0$ | √ | × | × | × |
| $t_1$ | × | √ | × | × |
| $t_2$ | √ | × | √ | × |
| $t_3$ | × | √ | × | √ |

Figures

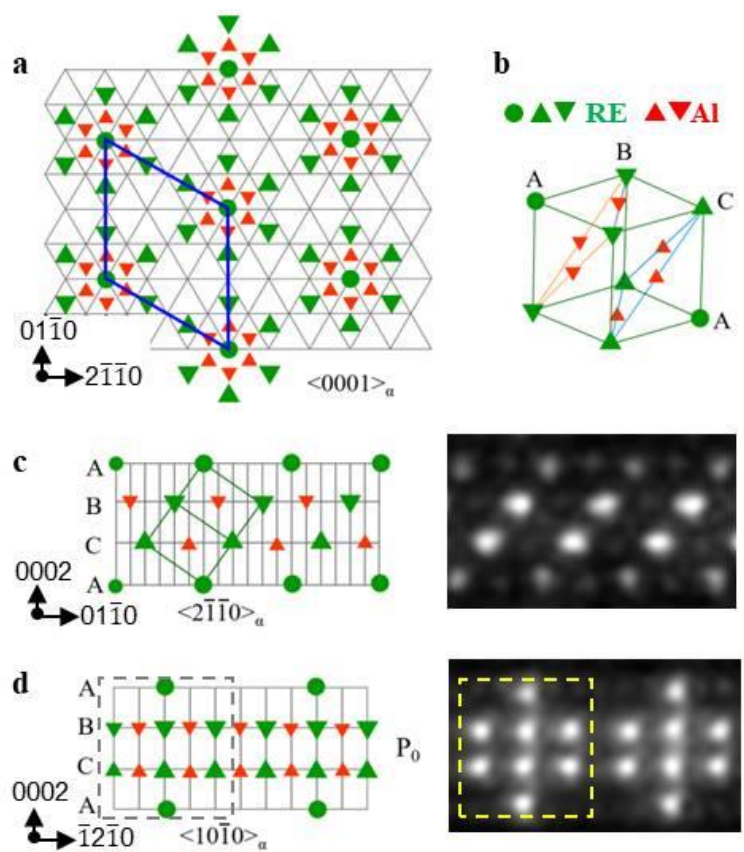

Fig. 1. Schematic diagram of ideal in-plane ordered clusters distributed in the fcc structural unit (SU) of LPSO structures. a) The four-layer height fcc SU with stacking sequence of "ABCA" viewed along $[0001]_\alpha$ in Mg ($\alpha$) matrix. Green color represents rare earth (RE) atoms, while red color is for Al atoms. The filled circles indicate the atoms in stacking layer A, downward-pointing triangle is for atoms in stacking layer B, and upward-pointing triangle is for atoms in stacking layer C. The nodes of grey grid show the Mg position in basal plane (A layer), and Mg atoms are omitted for simplicity. b) The $L1_2$ structure of the cluster in a). c) The $<2\bar{1}\bar{1}0>_\alpha$ view of the fcc SU in (a). A corresponding HAADF-STEM image is shown at the right side for reference. d) The $<10\bar{1}0>_\alpha$ view of the fcc SU. The $L1_2$ cluster in d) is marked by a dashed box. This ordered pattern is named as "$P_0$" type clusters for reference. (For interpretation of the references to color in this figure legend, the reader is referred to the web version of this article.)

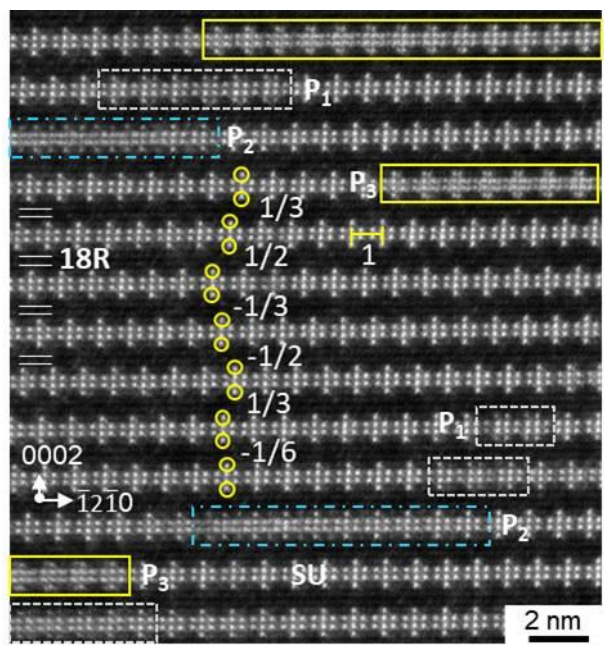

Fig. 2. An atomic image of the 18R LPSO structure observed along a $[10\bar{1}0]_\alpha$ zone axis by HAADF-STEM. Three new types of cluster patterns ($P_1$-$P_3$) in the SUs are indicated by different rectangles comparing to ideal $P_0$ pattern. The horizontal displacement between the clusters in neighboring SUs is referenced to the spacing ( = 1 ) between two $L1_2$ type clusters in a well ordered SU.

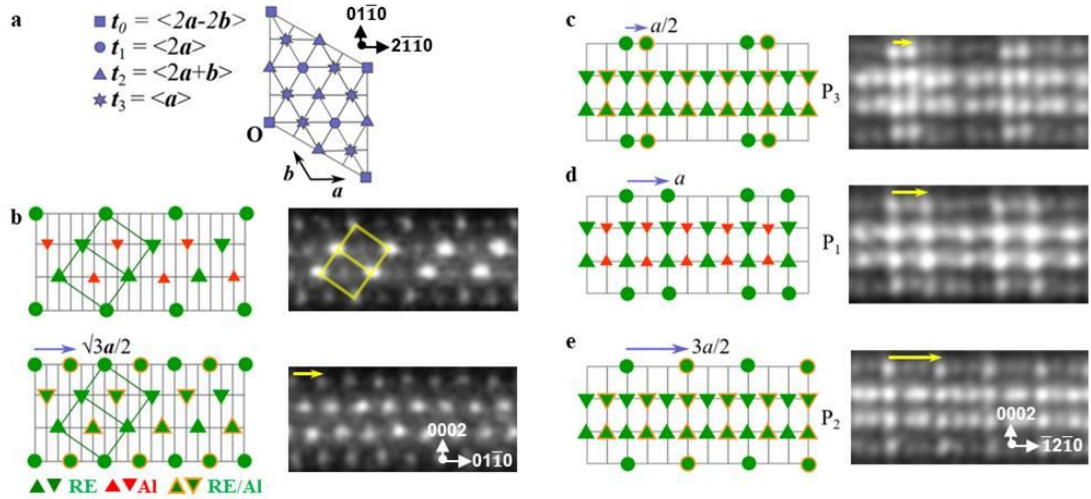

Fig. 3. Schematic diagrams of possible translational vectors for domain structures and different views of the superimposed domain structures. a) Four non-equivalent positions ($t_0$-$t_3$) in a rhombic lattice in A layer for generating a domain structure, where $t_0$ positions are the original translational positions for ordered clusters. Relative to the origin O, the positions for generating domains could be denoted as $t_0 = <2a\text{-}2b> = <2\bar{2}00>_\alpha$, $t_1 = <2a> = <4\bar{2}\bar{2}0>_\alpha/3$, $t_2 = <2a+b> = <10\bar{1}0>_\alpha$, $t_3 = <a> = <2\bar{1}\bar{1}0>_\alpha/3$, where $a$ and $b$ is the base vectors in the basal plane in Mg ($\alpha$) matrix. b) Two possible structures for domain structure $t_3$ viewed along different $<11\bar{2}0>_\alpha$, and a HAADF-STEM image is attached at right side for reference. c) The $<10\bar{1}0>_\alpha$ view of "$P_3$" type ordered pattern with a displacement of $a/2$ which is the projection length of some $t_3$. d) The $<10\bar{1}0>_\alpha$ view of "$P_1$" type ordered pattern with a projected displacement of $a$ for $t_1$ or some $t_3$. e) The $<10\bar{1}0>_\alpha$ view of "$P_2$" type ordered pattern with a projected displacement of $3a/2$ for $t_2$.

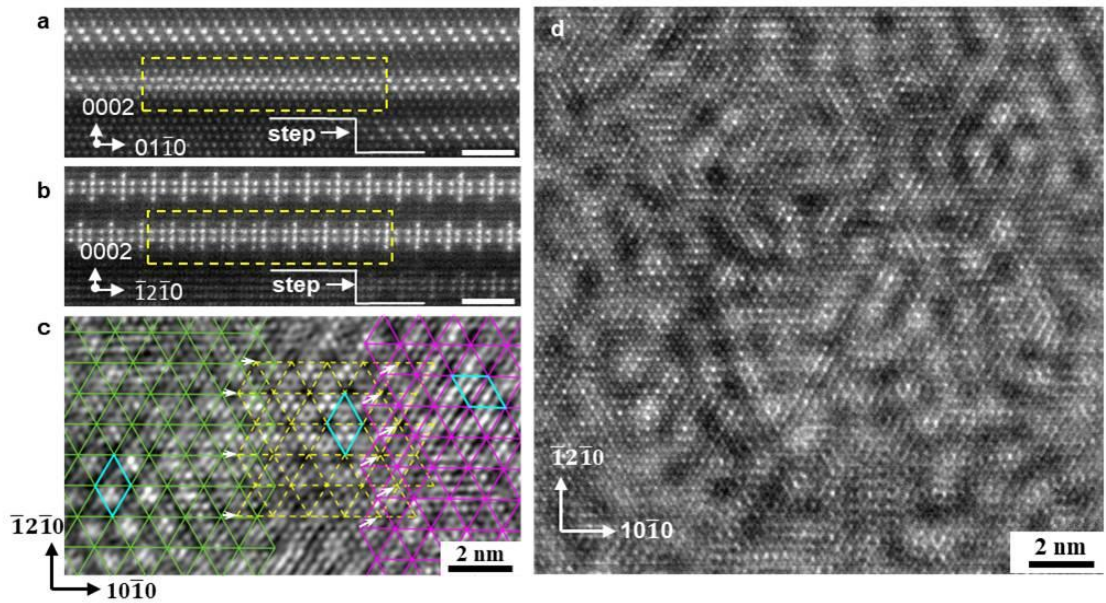

Fig. 4. Domain structures viewed at different zone axes by HAADF-STEM. a) $[2\bar{1}\bar{1}0]_\alpha$ zone axis view. b) Corresponding $[10\bar{1}0]_\alpha$ zone axis view of (a), where the '$P_1$' type clusters are marked by a yellow box, and the scale bar in (a-b) is 2 nm. c) Domain structures viewed along $[0001]_\alpha$. The possible distribution of clusters is indicated by grids with different color, and the mismatching between the grid lattices indicates the existence of domain structure. d) A comparing case of clusters in a structural unit in a deformed and aged $Mg_{97}Zn_1Gd_2$ alloy viewed along $[0001]_\alpha$.

Graphic abstract

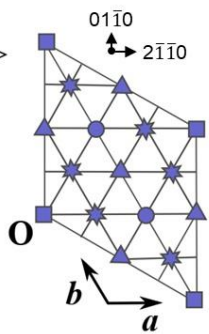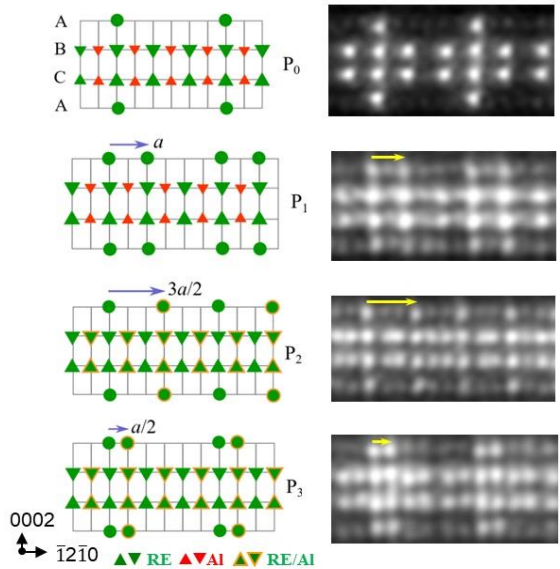

Different cluster patterns, viewed along $[10\bar{1}0]$, due to the superimposition of two domain structures specified by a relative translational vector $t$.

# Supplementary materials

## Domain structures and the transitional state of ordering in long-period stacking ordered (LPSO) structures in Mg-Al-Gd alloy

Xin-Fu Gu[a, b*], Tadashi Furuhara[b], Leng Chen[a], Ping Yang[a]

[a] School of Materials Science and Engineering, University of Science and Technology Beijing, Beijing, 100083, China

[b] Institute for Materials Research, Tohoku University, Sendai, 980-8577, Japan

*Corresponding author: Xin-Fu Gu, xinfugu@gmail.com

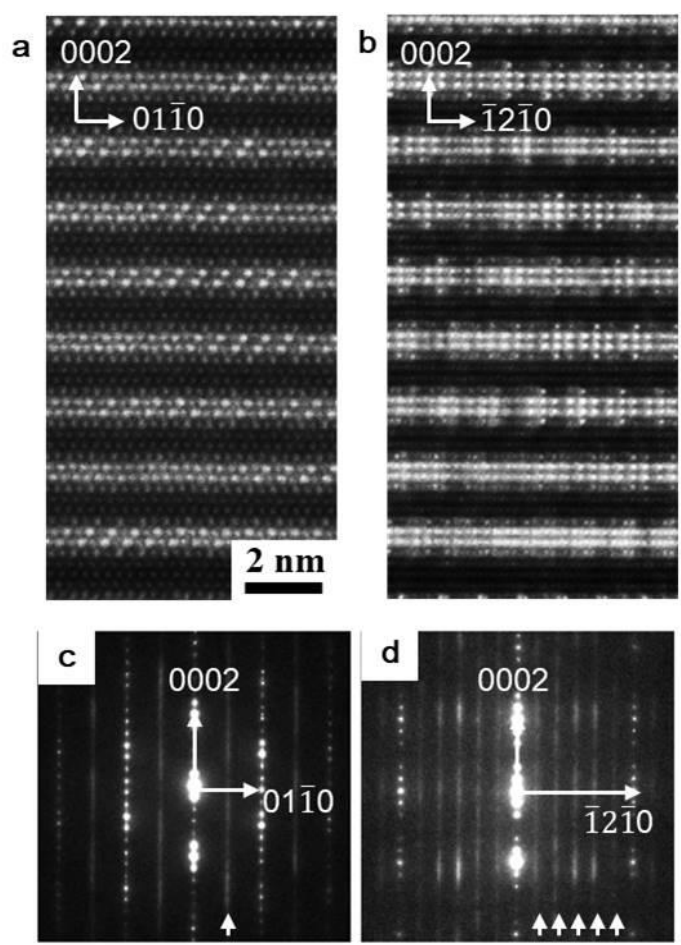

Fig. S1. 18R type of LPSO structures in as-casted Mg-Al-Gd alloy. HAAD-STEM image observed at the zone axis of a) $[2\bar{1}\bar{1}0]_\alpha$ and b) $[10\bar{1}0]_\alpha$. c) and d) are the corresponding diffraction patterns at two zone axes.

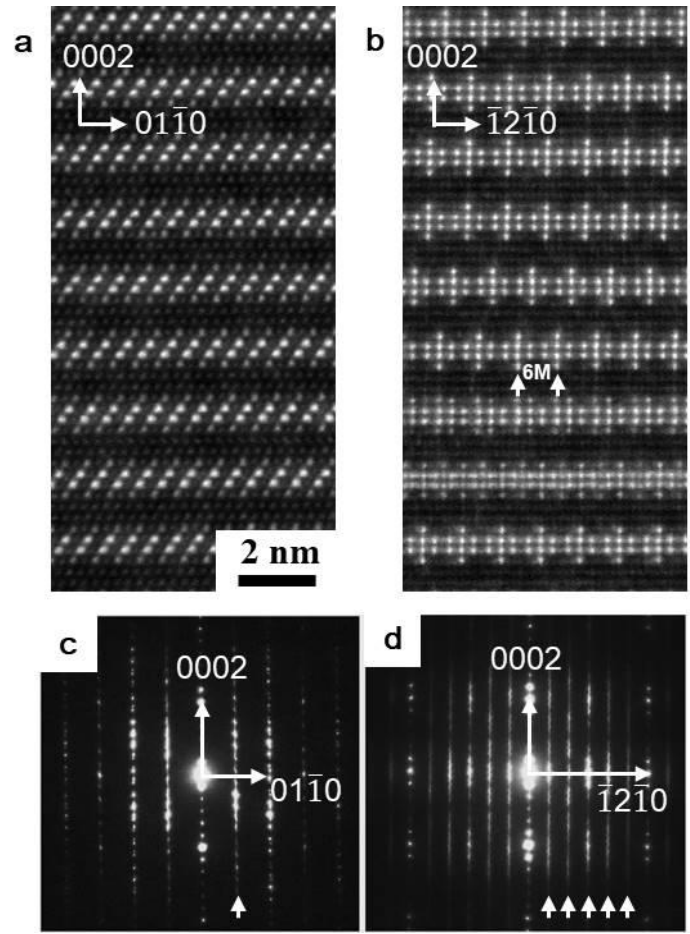

Fig. S2. 18R type of LPSO structures in as-casted Mg-Al-Gd alloy. HAAD-STEM image observed at the zone axis of a) $[2\bar{1}\bar{1}0]_\alpha$ and b) $[10\bar{1}0]_\alpha$. c) and d) are the corresponding diffraction patterns at two zone axes.